\documentstyle[natbib209,psfig,draft]{mn-nat}

\newcommand{\degree} {{^\circ}}
\newcommand{\RA}     {{\mathrm{RA}_{J2000}}}
\newcommand{\Dec}    {{\mathrm{Dec}_{J2000}}}
\newcommand{\OHMASER}    {{OH(1720~MHz)}}
\newcommand{\XRAYCOMPOSITE}    {{mixed morphology}}

\begin{document}

\title{327-MHz GMRT observations of one candidate and three Galactic
  Supernova Remnants}

\author[Bhatnagar Sanjay]
{Sanjay Bhatnagar\thanks{E-mail: sanjay@ncra.tifr.res.in} \\
National Centre for Radio Astrophysics (TIFR), \\
Pune University Campus, Post Bag No.3, Ganeshkhind, Pune 411 007,
India.\\ }
\date{1 Nov 2001}
\maketitle

\begin{abstract}
  Results from the 327-MHz Giant Meterwave Radio Telescope (GMRT)
  observations of four fields - three fields containing Galactic
  supernova remnants (SNRs) (G001.4$-$0.1, G003.8$+$0.3, and
  G356.3$-$1.5) and one field containing a candidate SNR
  (G004.2$+$0.0), are reported in this paper.  These fields were
  selected from the 843-MHz survey conducted by \citet{Gray-III} using
  the Molonglo Synthesis Telescope (MOST).  All the three SNRs are
  detected in the GMRT images.  Significant amount of thermal emission
  is seen at the location of the candidate SNR, and the GMRT image
  shows a discrete source of emission which is consistent with it
  being a flat spectrum thermal source.  An incomplete arc of emission
  with a compact central source detected in one of the fields, is
  coincident with the extended \OHMASER\ emission reported earlier.
  Possible implications of this morphology and correlation with the
  \OHMASER\ emission are also discussed.
\end{abstract}


\section{INTRODUCTION}

Galactic objects are usually classified on the basis of their
morphology and spectra.  In a supernova explosion, the interaction of
the blast wave with the interstellar medium produce extended objects
in the sky which emit non-thermal radiation.  Extended emission
showing one of the usual morphologies of Galactic Supernova Remnants
(SNRs) and a negative spectral index $\alpha$ ($S \propto \nu^\alpha$)
(indicative of non-thermal emission) have been the primary signatures
used to identify SNRs.  In this context, high resolution, sensitive
imaging in the Galactic plane at low frequencies offers several
advantages.  Synchrotron emission from SNRs becomes progressively
stronger relative to thermal emission, which dominates at higher
frequencies.  Hence, although high frequency observations often
provide the required resolution, they suffer from contaminating
thermal emission, particularly in complicated regions like the
Galactic plane.  On the other hand, higher resolution observations at
frequencies below $\sim1$~GHz can effectively separate thermal and
non-thermal emission as well as reliably decipher the morphologies of
the objects in the field.

Recently, synthesis telescopes like the Giant Meterwave Radio
Telescope (GMRT) and the Molonglo Synthesis Telescope (MOST) have been
used to map parts of the Galactic plane.  \citet{Gray-III} used the
MOST for imaging in the Galactic plane at a resolution of $\sim 90
\times 43.5$~arcsec$^2$ at 843~MHz.  These
observations were sensitive to angular scales up to $30$~arcmin and 17
candidate Supernova remnants (SNRs) were identified from this survey,
many of which are large angular size objects ($>5-10$~arcmin).  Many
of the fields however suffer severely from the grating response and
from confusing thermal emission from nearby strong sources.

\citet{YoursTruely2000} used the GMRT \citep{GMRT} at 327~MHz to
confirm three SNRs selected from the survey done by
\citet{DUNCAN_SNRFrom2.4GHzSurvey} at 2.4 GHz using the Parkes 64-m
dish.  The GMRT at 327~MHz provides a resolution of $\sim20$~arcsec
for observations in the southern sky and is sensitive to spatial
scales of up to 30~arcmin.  At this frequency, thermal emission from
typical HII regions is weak while the emission from SNRs remain
relatively strong.  The relatively smaller field-of-view ($\sim
1^\circ.4$ at 327~MHz) of the GMRT offers a further advantage in terms
of attenuating confusing emission farther away.  A relatively better
imaging performance of the GMRT compared to that of the MOST allows
higher dynamic range mapping of complicated fields like the ones
imaged by \citet{Gray-III} and hence more reliable maps at a higher
resolution.  Thus the combination of high angular resolution,
sensitivity to large scale structures and comparatively higher
sensitivity that the GMRT provides at these low frequencies makes it a
good instrument for studies of Galactic SNRs.

I present here the results of 327~MHz GMRT observations of three
objects (namely, G001.4$-$0.1, G003.8$+$0.3, and G356.3$-$1.5) which
have been included in the Galactic SNR Catalogue \citep{SNRCAT} based
on 843-MHz \citep{Gray-III} observations alone, and one candidate SNR
(G004.2$+$0.0).  These fields were selected from the 843-MHz survey by
\citet{Gray-III}.

\section{Observations}

The observations were made with the GMRT \citep{GMRT} during
$1999-2000$.  During this period, the GMRT correlator as well as the
control software were in a state of being debugged.  The maximum
baseline available therefore changed from observation to observation,
due to non-availability of certain antennas.

All observations were made using the single side-band GMRT correlator
which computes, as of now, only the co-polar visibilities (only
signals of same polarization from different antennas are multiplied).
This correlator handles a maximum input bandwidth of 16~MHz and
provides 128 channels across the band.  The full 16~MHz band centered
around 327~MHz was used, resulting in 128 frequency channels of width
$\approx 125$~kHz per baseline.  The procedure for data
flagging/editing, calibration and image restoration was similar to
that described in \citet{YoursTruely2000}.

Each field was observed for $\sim 4$~hours.  Out of the total of the
30 antennas of the GMRT, 12 are within an area of size
$\approx1$~km$\times$1~km, referred to as the Central Square.  These
provide the short spacing uv-coverage crucial for mapping extended
emission.  The shortest spacing from which data were included in these
observations was $\sim100~\lambda$ making these observations sensitive
to a maximum angular scale of about $\sim25$~arcmin.  All the sources
reported here are significantly smaller than this limit and hence do
not suffer from the problem of missing flux.  The compact 28-Jy source
1830$-$36 was observed once every 30~min as the phase calibrator.  The
observations were amplitude calibrated using 3C48 which was assumed to
have a flux density of 42.7 Jy.  The phase calibrator (1830$-$36) was
also used for bandpass calibration to produce the multi-channel
continuum maps.  Most of the data reduction was done using the
{\tt AIPS} package of the National Radio Astronomy Observatory (NRAO),
USA.  After flagging radio frequency interference (RFI) affected and
otherwise bad data, a total bandwidth of $\sim4$~MHz was finally
used for imaging.  Since all the fields are complex with strong
sources distributed all over the primary beam, the data were kept in
the multi-channel format throughout the mapping process, in order to
minimize the effect of bandwidth smearing of source away from the
phase centre.  With a field of view of $\sim1\degree.4$ of the GMRT
antennas, the distortions of sources away from the phase centre due to
the w-term become important \citep{TIM_N_RICK_1992}.  Hence, all
images were made using the multi-facet 3D imaging algorithm
implemented in the {\tt IMAGR} task of the {\tt AIPS} package.

\section{Results}

The parameters of the observations and the observed and derived
physical parameters of the SNRs in the fields are listed in
Table~\ref{SNR_PARAMS}.  The image of one of the SNR (G356.3$-$1.5)
from Gray's survey was severely affected by the grating responses due
to nearby strong sources.  A barrel shaped SNR is detected in the GMRT
image of this field.  A weak shell-type SNR is detected in the second
field (G003.8$+$0.3) while no significant emission feature is detected
in the field containing G004.2$+$0.0.  I believe that this latter
source is not an SNR.  In the field of G001.4$-$0.1, besides other
well known extended sources in the Galactic Centre region
\citep{GC@327MHz}, a clear partial arc with a compact central source,
set almost at the geometric centre of the arc is detected.  This arc
of possibly non-thermal emission coincides fairly well with the
extended \OHMASER\ emission reported by \citet{OH_SNR_EMISSION}.  This
morphology suggests that this object may belong to the newly
recognized class of composite SNRs with significant X-ray emission
\citep{10^51Ergs}.

Detailed results of the individual objects are given below.

\begin{table*}
\begin{center}
\caption{Observed and derived parameters of the SNRs.  Type code 'S'
  implies shell-type, 'B' implies barrel-type SNR and 'N' implies that
  the object is not an SNR}.
\label{SNR_PARAMS}
\vskip 1cm
\begin{tabular}{|l|c|c|c|c|c|c|c|c|}
\hline
Name      & RA      &Dec    &$S_{327MHz}$&Resolution  &RMS   &Size       &Type&$\alpha$ \\
          &(J2000)  &(J2000)&(Jy)        &(arcsec$^2$)&(mJy/beam)&(arcmin)&    &$(S\propto\nu^\alpha$)\\
\hline
G001.4$-$0.1&$17^h49^m39^s$&$-27\degree45\arcmin$&$4.2\pm0.5$&$162\times144$&10&8 &S&$?$\\
G003.8$+$0.3&$17^h53^m02^s$&$-25\degree24\arcmin$&$6.0\pm0.4$&$27\times17$&3&18&S  &$-0.6\pm0.1$\\
G004.2$+$0.0&$17^h55^m22^s$&$-25\degree15\arcmin$&$0.1\pm0.007$&$144\times78$&7&-&N&-\\
G356.3$-$1.5&$17^h42^m40^s$&$-32\degree52\arcmin$&$5.7\pm0.2$&$102\times48$&3&15&S/B&$>-0.7\pm0.1$\\
\hline
\end{tabular}
\end{center}
\end{table*}

\subsection{G001.4$-$0.1}

This SNR lies at the very edge of the 327-MHz wide-field VLA image of
the Galactic centre region and just north-east of the HII region Sgr~D
\citep{GC@327MHz}.  Extended low surface brightness emission is seen
in this wide-field image corresponding to this SNR.  However the image
quality for this region is too poor (due to primary beam attenuation)
to be able to measure the flux density or decipher the morphology.

\begin{figure}
\begin{center}
\psfig{figure=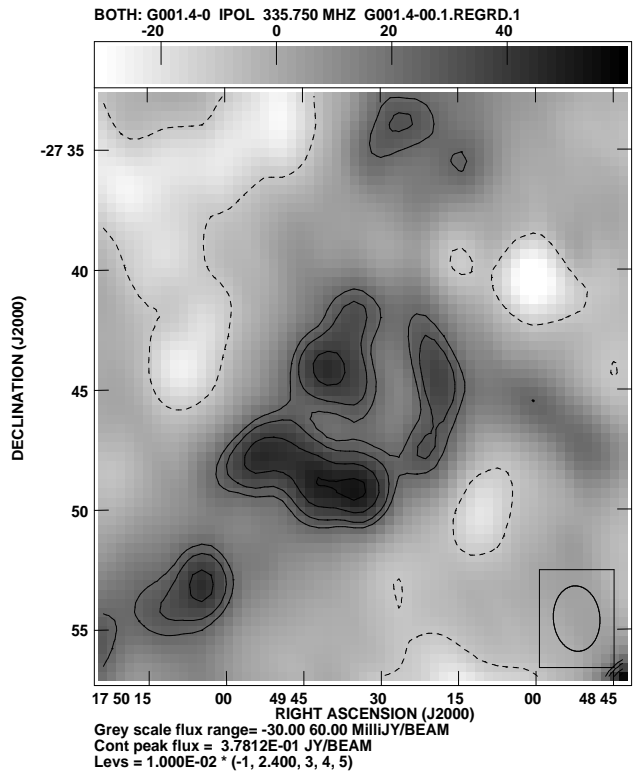,width=3.25in,height=4in}
\caption[GMRT 327-MHz image of G001.4$-$0.1]{327-MHz image of
  G001.4$-$0.1 using GMRT.  The RMS noise in the image is $\sim
  10$~mJy/beam and resolution of $\sim 2.7 \times 1.9$~arcmin$^2$ at a
  P.A. of $5\degree$.  A partial shell with compact source located
  almost at the centre is clearly detected.}
\label{G1.4-0.0@327MHz}
\end{center}
\end{figure}

The GMRT 327-MHz image of G001.4$-$0.1 (Figure~\ref{G1.4-0.0@327MHz})
clearly shows a partial arc of $\sim 8$~arcmin diameter.  A compact
source, almost at the geometric centre of the arc is also clearly
visible.  This SNR has a morphology very similar to that of the nearby
composite SNR G000.9$+$0.1 which has a flat spectrum, X-ray emitting
compact central source \citep{G0.9-0.1_XRAY,Sidoli_2000}.  The 327-MHz
flux density measured from our observation was found to be
$4.2\pm0.5$~Jy.

The MOST image at 843 MHz suffers from artifacts due to the grating
response of Sgr~A and the flux density of $\sim2$~Jy at 843~MHz is
stated to be only a ``tentative'' figure.  This image shows a
relatively featureless source of emission and the ``complete shell''
reported by \citet{Gray-III} is difficult to decipher.
\citet{G1.4-0.0@1616MHz} reported ``an arc of incomplete shell'' of
diameter $\sim7$~arcmin from the 1616-MHz image of this region.  The
incomplete arc seen in the GMRT image agrees well with the incomplete
arc seen in the 1616-MHz image.

A local maximum of emission is also clearly detected in the
Parkes-MIT-NRAO (PMN) survey image at 4.85~GHz \citep{PMN2} which has
a resolution of about $\sim5$~arcmin.  The flux density measured from
this image is $\sim 4.2$~Jy.  However the extent of emission is much
larger than that seen at 327~MHz and hence this value probably
includes a substantial contribution from the larger scale background
emission in this region.

This source was also the target of search for the \OHMASER\ maser
emission by \citet{OH_SNR_EMISSION}.  In the VLA A-array observation,
they detected a maser spot, coincident with the western edge of the
arc.  Their VLA D-array observations detects an extended arc of
\OHMASER\ emission, almost coincident with the arc seen in the radio
continuum images mentioned above.

\OHMASER\ maser emission is positionally and kinematically associated
with SNRs \citep{Claussen_1997}.  Based on the morphology and
association of \OHMASER\ emission, we propose that this in a
shell-type SNR in the Galactic plane.  There is evidence for possible
non-thermal nature of emission based on the 843- and 327-MHz flux
densities.  However given the uncertainty in the 843-MHz flux
densities, this needs to be established by measurements at other
frequencies. 

\subsection{G003.8$+$0.3}

\begin{figure}
\begin{center}
\psfig{figure=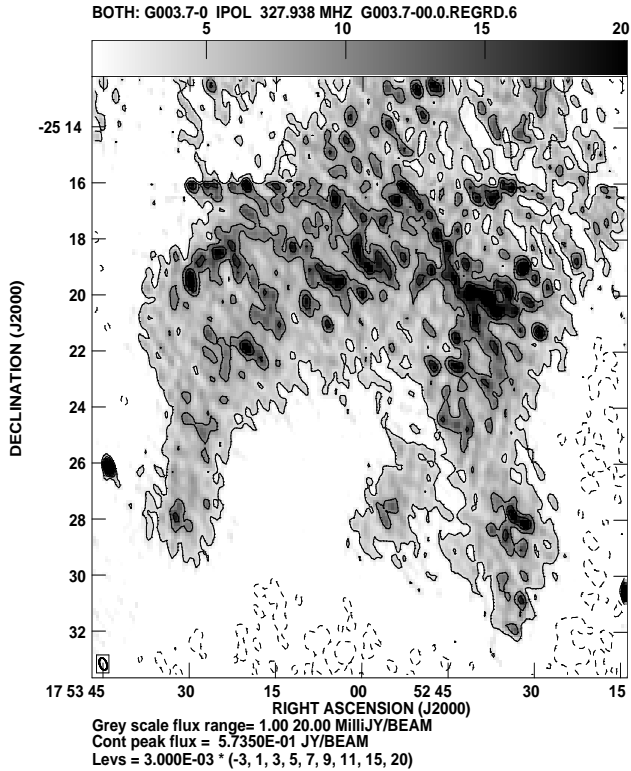,width=3.25in,height=4in}
\caption[A 3D inverted GMRT 327-MHz image of G003.8$+$0.3]{GMRT image
  of G003.8$+$0.3 at 327~MHz.  Shell morphology of G003.8$+$0.3 is
  clearly visible.  The negative immediately south of this source
  (seen even in the 843-MHz image of \citet{Gray-III}.) may account
  for the apparent incompleteness of the shell.  The RMS noise in the
  images is $\sim 3$~mJy/beam and a resolution of $\sim
  27\times17$~arcsec$^2$ at a P.A. of $23\degree$.}
\label{G3.8+0.3@327MHz}
\end{center}
\end{figure}

\citet{Gray-III} reported G003.8$+$0.3 as a ``fairly week, incomplete
ring structure most perfectly centered on a slightly extended
source''.  The morphology of this object in the GMRT 327-MHz image
(Figure~\ref{G3.8+0.3@327MHz}), is consistent with the 843-MHz MOST
image.  The IRAS 60$\mu$m image
does not show any significant emission at this position.  The northern
rim of the ring is significantly brighter and more extended making it
difficult to define the center of the ring structure.  The ``central
source'', located at $\RA= 17^h52^m 54^s$, $\Dec=-25^\circ28\arcmin$
in the image, is close to the center defined by the inner edge of the
ring structure, but not close to the center defined by the outer
edges.  The bridge of emission connecting the central source and the
ring is not so clearly seen at 327~MHz.  However there is a rather
deep negative region immediately south of this SNR and it is possible
that the shell appears partial due to the presence of the negative
region in this image.

The integrated flux density of this region at 327~MHz is
$6.0\pm0.4$~Jy.  
The diameter of the ring structure
(including the northern extension) is $\approx 18$~arcmin and center
is at $\RA=17^h53^m02^s$, $\Dec=-25^\circ24\arcmin$.
The flux density reported by Gray at 843~MHz is 3.5~Jy giving a
spectral index of $-0.6\pm0.1$ (here and elsewhere in this paper, the
quoted RMS noise of $\sim 5$mJy/beam of the MOST survey \citep{Gray-I}
was used to compute the random error for the 843-MHz flux density).
Based on the morphology and the evidence for non-thermal emission, we
propose that this source is a faint Galactic SNR.

The ``central source'' is detected as a point source at 1400~MHz in
the NRAO VLA Sky Survey or NVSS \citep{NVSS} with a flux density of
$15.1$~mJy.  Although it is a weak source in the 327-MHz image and
flux density is barely at the $2\sigma$ level, it is nonetheless
stronger than $15.1$~mJy at 1400 MHz, indicating that it may be
non-thermal.  With a large error in the 327-MHz flux density, it is
difficult to determine an accurate enough spectral index.  There is no
pulsar in the vicinity of this source and the nature of this central
source remains unknown.  As noted by \citet{Gray-III}, it could be a
chance superposition of an unrelated background extra-galactic source.
A connecting bridge of emission seen in the 843-MHz image is however
not as clearly seen at 327~MHz.

\begin{figure}
\begin{center}
\psfig{figure=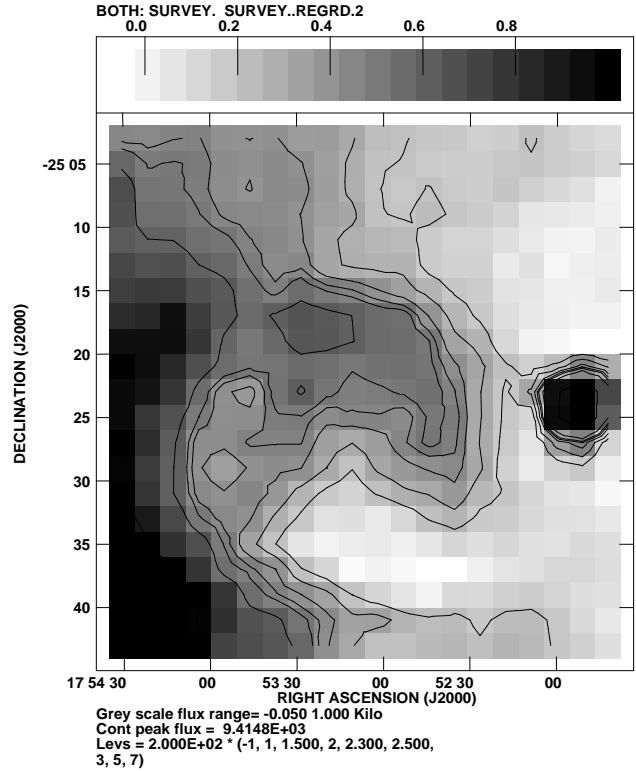,width=3.25in,height=4in}
\caption{Image of G003.8$+$0.3 at 11~cm from the 100-m Effelsberg single
  dish survey.  An arc of emission similar to that seen at lower
  frequencies is clearly visible.  The strong compact source towards
  the west is the L-Band VLA Calibrator J$1751-253$
  ($17^h51^m51.3^s,-25^\circ 24'00.06"$).}
\label{G3.8+0.3@11cm}
\end{center}
\end{figure}

Image of this SNR from the radio continuum survey of the Galactic
plane at 11~cm using the 100-m Effelsberg telescope \citep{11cmGP} is
shown in Figure~\ref{G3.8+0.3@11cm}.  Here too, a partial shell of
emission is visible, which matches well with the morphology of this
SNR seen at 843 and 327~MHz.  Significant amount of linear
polarization is also detected in the 11~cm polarized images.

\subsection{G004.2$+$0.0}

\begin{figure}
\begin{center}
\psfig{figure=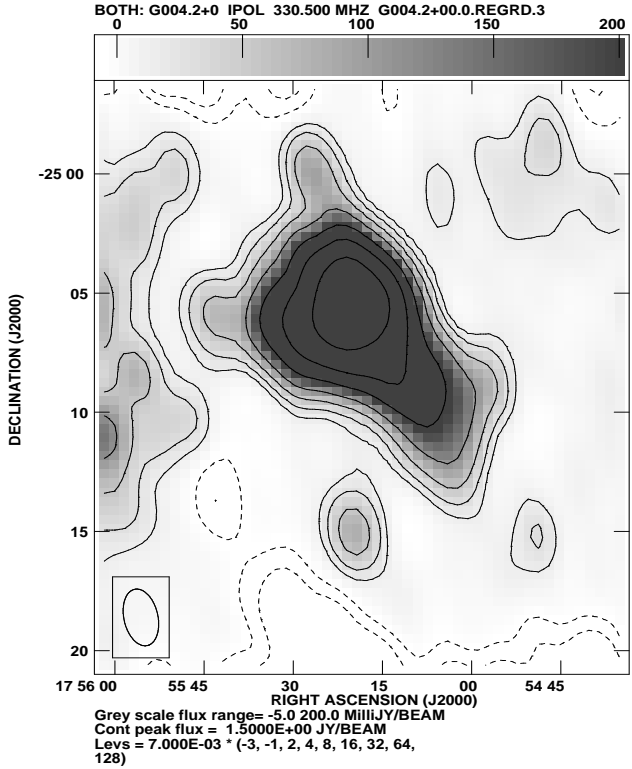,width=3.25in,height=4in}
\caption[GMRT 327-MHz image of extended source north of
G004.2$+$0.0]{GMRT 327-MHz image showing an extended source located
  just north of the candidate SNR G004.2$+$0.0.  The RMS noise in this
  images is $\sim 7$~mJy/beam and the resolution is $\sim 2.4 \times
  1.3$~arcmin$^2$ at a P.A. of $10.6\degree$.  The known HII region
  G004.4$+$0.1, coincides with this source at
  $\RA=17^h55^m26^s,\Dec=-25\degree05\arcmin08\arcsec$.  It's
  south-western tail is also visible in Grey's image at 843~MHz.}
\label{G4.2-0.0@327MHz}
\end{center}
\end{figure}

\begin{figure}
\begin{center}
\psfig{figure=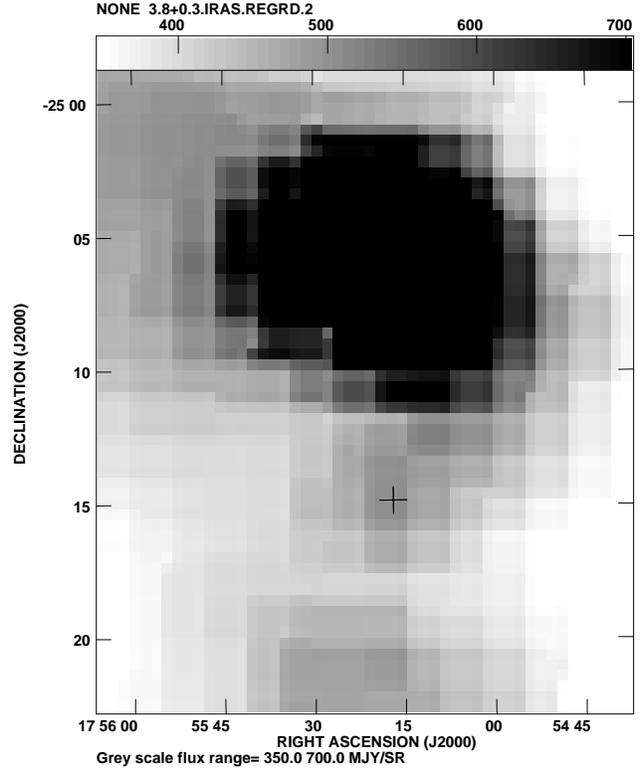,width=3.25in,height=4in}
\caption[IRAS $60\mu m$ image of G004.2$+$0.0]{The IRAS $60\mu$m image
  of the region containing G004.2$+$0.0.  The HII region G004.4$+$0.1
  seen in the GMRT 327-MHz image is the strongest source in this
  image.  A faint source at the position of G004.2$+$0.0, indicated by
  a cross, is also detected in this IRAS image.}
\label{G4.2+0.0.IRAS}
\end{center}
\end{figure}

This source is the smallest diameter candidate SNR (size 3.5~arcmin)
reported by \citet{Gray-III}.  He reported the location of this object
as $\RA=17^h55^m17^s, \Dec=-25\degree14\arcmin51\arcsec$.  The total
843-MHz flux density was reported to be $200$~mJy.  However this
object sits in a negative bowl and the measured value after tentative
correction for the negative bowl is in the range of $100-300$~mJy
(Gaensler, private communication).
A high resolution image of this field was made using the GMRT to look
for the shell-type structure at 327 MHz.  There is a hint of a compact
source in this image at $\RA=17^h55^m 22^s, \Dec= -25^\circ15\arcmin
01\arcsec$, but barely at $2\sigma$ level.  No shell-type structure
was detected at the level of $\sim 5$~mJy/beam with a resolution of
$\sim 15$~arcsec.  The low resolution image, shown in
Figure~\ref{G4.2-0.0@327MHz} shows a $\sim 100$~mJy object at the
location of this source.  There is probably a compact source in the
NVSS image at this location, but again barely at $1-1.5\sigma$ level
(which cannot be treated as a detection).  The $60\mu m$ IRAS image of
this source (Figure~\ref{G4.2+0.0.IRAS}) also shows significant
extended emission at the location of this source (indicated by a cross
in the figure), which appears to be associated with the HII region in
the north, indicating that this may be a thermal source.  This source,
based on the available radio flux densities is therefore consistent
with it being a flat spectrum thermal source and may not be an SNR.

The dominant extended source in the image shown in
Figure~\ref{G4.2-0.0@327MHz} is a known HII region, G004.4$+$0.1 located
at $\RA=17^h55^m26^s, \Dec=-25\degree05\arcmin08\arcsec$
\citep{HII_REGIONS@4GHz}.  A compact core surrounded by a halo of
lower surface brightness is clearly visible in this image and this
core-halo morphology is suggestive of this being a compact HII region
\citep{UCHII_MORPHOLOGIES}. In combination with high resolution images
at other frequencies, these data can provide information about the
physical conditions in this HII region which will be reported
elsewhere.

\begin{figure}
\begin{center}
\psfig{figure=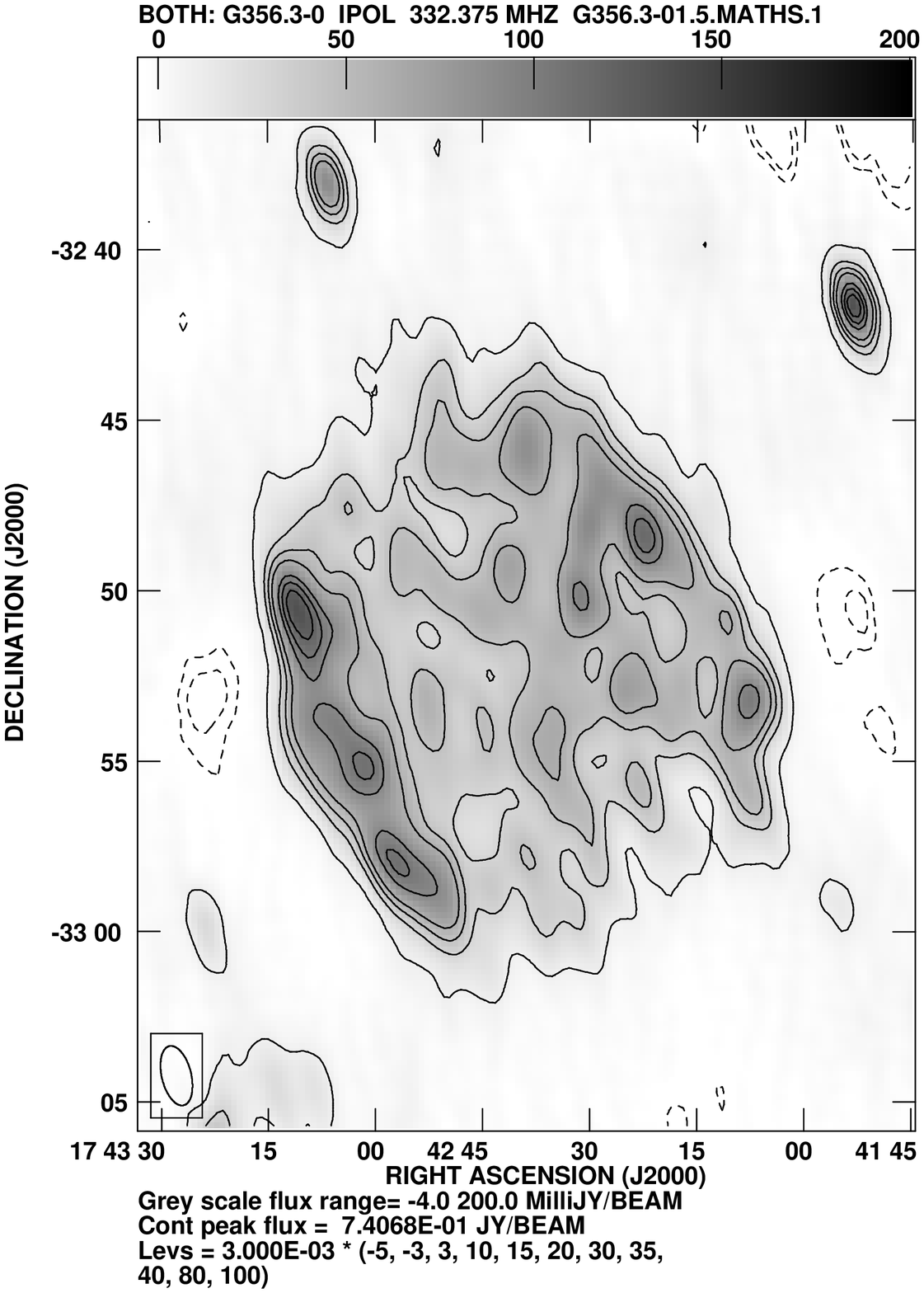,width=3.25in,height=4in}
\caption[GMRT 327-MHz image of G356.3$-$1.5]{327-MHz image of
  G356.3$-$1.5 using GMRT.  The box shaped morphology of the object is
  apparent.  The emission filling the center of the object is not
  detected in the 843-MHz MOST image of Gray.  The RMS noise in the
  image is $\sim 3$~mJy/beam and a resolution of $\sim 1.7\times
  0.8$~arcmin$^2$ at a P.A. of $14\degree$.}
\label{G356.3-1.5@327MHz}
\end{center}
\end{figure}

\subsection{G356.3$-$1.5}

This SNR was termed by \citet{Gray-III} as a 'classic barrel' based on
the 843-MHz image.  The GMRT 327-MHz image, shown in
Figure~\ref{G356.3-1.5@327MHz} confirms the basic structure seen in
the 843-MHz image where the two edges are relatively brightened
compared to the central region.  However, at 327~MHz, the center is
also filled with significant emission.

The integrated flux density measured at 327~MHz from the GMRT image is
$5.7\pm0.2$~Jy.  The RMS noise in the image in the vicinity of this
object is about 3~mJy/beam.  The integrated flux density in the image
at 843~MHz is reported to be 2.8~Jy.  This implies a spectral index of
$-0.7\pm0.1$ between 843 and 332~MHz.  The 843-MHz flux density
however is likely to be an underestimate (due to the subtraction of
smooth component to remove the grating response) and hence the
resulting spectral index a lower limit.

\section{Discussion}

A faint local maximum surrounded by extended emission is detected at
the location of G001.4$-$0.1 in the 11~cm radio continuum survey of the
Galactic plane using the 100-m Effelsberg telescope \citep{11cmGP}.
However the angular resolution in this survey ($\sim 5$~arcmin) is not
enough to decipher the shell morphology of this SNR.  G003.8$+$0.3 is
however clearly detected as a shell like structure in this survey and
it is somewhat surprising that it was not even classified as a
candidate SNR prior to the 843-MHz observations of this field.

The emission in the 843-MHz image of G356.3$-$1.5 drops to almost
noise level between the rims.  Apart from this, the general morphology
is very similar to that seen in the 327-MHz image.  The 843-MHz image
suffered from the grating response of the nearby source G357.7$-$0.1
which ran right across this source.  An attempt was made to remove
this, and as reported by \citet{Gray-III}, this subtraction was not
entirely successful.  It therefore appears that the clear minimum
between the rims which appears in the 843-MHz image, may actually be
an artifact and not real.  This is also supported by the fact that
compared to the relatively uniform emission filling the centre at 327
MHz, the central emission in the final image at 843 MHz is
significantly lower than the emission from the western rim.

The surface brightness at 1~GHz for G003.8$+$0.3 and G356.3$-$1.5 is
$1.43\times10^{-21}$ and $>1.75\times10^{-21}$
W~m$^{-2}$~Hz$^{-1}$~sr$^{-1}$ respectively.  In the absence of any
independent estimate of the distance to these SNRs the current data
cannot throw any light on the $\Sigma-D$ relation except to note that
the surface brightness of these SNRs is on the lower side.  The
distance calibrators show a larger scatter on the $\Sigma-D$ plane in
this range of surface brightness \citep{NEWSigma_D}.  A reliable
independent distance estimate to these SNRs in the future could be
useful for improving the calibration of the $\Sigma-D$ relation
itself.

The morphology of some of the SNRs have long been suspected to be
shaped by their interaction with nearby molecular clouds.  Recently it
has been argued that the \OHMASER\ emission is a good tracer of this
interaction \citep{Frail_Goss_1994}.  The \OHMASER\ maser emission is
distinguished from the OH maser emission at 1665, 1667 and 1612 MHz by
the former being positionally and kinematically associated with SNRs
\citep{Claussen_1997} while the latter are associated with HII
regions.  Both theoretical and observational evidence
\citep{Reach_Rho_1998,Reach_Rho_1999,Frail_Mitchell_1998} suggests
that the \OHMASER\ masers are associated with the C-type shocks and
are collisionally pumped in molecular clouds at temperatures and
density of $50-125$~K and $10^5-10^6$~cm$^{-3}$ respectively
\citep[and references therein]{Lockett1999}.  OH masers at 1665, 1667
and 1612 MHz cannot be produced under these physical conditions and
the absence of these lines along with \OHMASER\ favours this
interpretation.  The measurements of the post shock density and
temperature for IC443 \citep{vanDishoeck_1993}, W28, W44 and 3C391
\citep{Frail_Mitchell_1998} are in excellent agreement with these
theoretical predictions.  A solution to the problem of producing OH,
which is not directly formed by shocks, is proposed by
\citet{Wardle1999_2}.  They suggest that the molecular cloud is
irradiated by the X-rays produced by the hot gas in the interior of
the SNR.  This leads to photo-dissociation of the H$_2$O molecules,
which is produced by the shock wave in copious amounts, behind the
C-type shock resulting into the required enhancement of OH just behind
the C-type shock.  The observed association of \OHMASER\ with
\XRAYCOMPOSITE\ SNRs (i.e., shell-type radio morphology with a centre
filled X-ray morphology) and strong correlation between the morphology
of the molecular gas and synchrotron emitting relativistic gas
\citep{Frail_Mitchell_1998} offers observational evidence for the
hypothesis that the \OHMASER\ maser originates in the post-shock gas,
heated by the SNR shock passing through dense molecular clouds.

The detection of \OHMASER\ maser spot towards G001.4$-$0.1
\citep{OH_SNR_EMISSION} suggests that the SNR is driving a C-type
shock in a molecular cloud.  The clear arc of emission in the radio
continuum image at 327~MHz is morphologically similar to the extended
\OHMASER\ emission towards this source lending further support to this
model.  The extended \OHMASER\ emission may then be tracing the SNR
shock front interaction with the molecular gas and may be associated
with the compact \OHMASER.  The absence of any such cloud on the north
eastern side of this SNR can explain the incomplete arc seen in radio
continuum as well as in \OHMASER\ emission.  Also, similarity of the
radio morphology with that of G000.9$+$0.1, which is known to be a
\XRAYCOMPOSITE\ SNR \citep{G0.9-0.1_XRAY,Sidoli_2000} with an
associated \OHMASER\ maser emission \citep{OH_SNR_GREEN_1997}, also
suggests that this SNR also may be of the \XRAYCOMPOSITE\ class.
Alternatively, the extended \OHMASER\ emission may be the widespread
\OHMASER\ emission from the foreground gas with the background
emission coming from the SNR
\citep{OH_Haynes_Caswell_1977,OH_Turner_1982}.  A detailed analysis of
the velocity structure of the extended \OHMASER\ emission is required
to conclusively determine the nature of this emission and if it is
kinematically associated with the compact emission.  Determination of
the radio spectra of the shell and the central source and X-ray
observations of this SNR will be the useful future observations to
determine the nature of this SNR.

\citet{Caswell_1977PASAu} pointed out that many SNRs are brighter
towards the Galactic plane.  Attempt to explain this as a being due
to a global density gradient of the interstellar medium (ISM) has
been shown to be statistically not significant \citet{GREEN84} and is
now believed to be due more to local effects rather than a global
effect.  In this context, it is significant to note that G003.8$+$0.3
and G356.3$-$1.5 are both brighter {\it away} from the Galactic plane.

\section{Conclusions}

In conclusion, we confirm G001.4$-$0.1, G003.8$+$0.3 and G356.3$-$1.5
to be Galactic SNRs based on the morphology and non-thermal nature of
emission from these objects.  These SNRs were reported in the list of
candidate SNRs by \citet{Gray-III} and have since then been include in
the latest catalogue of Galactic SNRs \citep{SNRCAT} based on the
morphological evidence alone.  These observations add the new
information of the non-thermal nature of emission from these objects
which firmly establishes them as Galactic SNRs.  Significant thermal
emission is seen at the location of another small diameter candidate
SNR G004.2$+$0.0 reported by \citet{Gray-III}.  The radio flux
densities of this object are consistent with it being a thermal
source.  Morphology of G001.4$-$0.1, coincidence of a compact
\OHMASER\ spot with the western arc and morphologically similar
extended \OHMASER\ emission towards this sources is suggestive of an
interaction with a nearby molecular cloud.

\section{ACKNOWLEDGMENTS}

It is a pleasure to thank Namir Kassim, Gopal-Krishna, and Bryan
Gaensler for their very useful comments and suggestions.

I thank the staff of the GMRT that made these observations
possible. GMRT is run by the National Centre for Radio Astrophysics of
the Tata Institute of Fundamental Research.

This research has made use of NASA's Astrophysics Data System Abstract
Service.  Almost all of this work was done using computers running
the GNU/Linux operating system and it is a pleasure to thank all the
numerous contributors to this software.  The AIPS package developed at
National Radio Astronomy Observatory, USA, was extensively used for
this work and I wish to thank the authors of this software.

\bibliography{mnras-abv,mb376}
\bibliographystyle{mn}

\end{document}